\documentclass[10pt,a4paper,onecolumn]{article}
\usepackage{marginnote}
\usepackage{graphicx}
\usepackage{xcolor}
\usepackage{authblk,etoolbox}
\usepackage{titlesec}
\usepackage{calc}
\usepackage{tikz}
\usepackage{hyperref}
\hypersetup{colorlinks,breaklinks=true,
            urlcolor=[rgb]{0.0, 0.5, 1.0},
            linkcolor=[rgb]{0.0, 0.5, 1.0}}
\usepackage{caption}
\usepackage{tcolorbox}
\usepackage{amssymb,amsmath}
\usepackage{ifxetex,ifluatex}
\usepackage{seqsplit}
\usepackage{xstring}

\usepackage{float}
\let\origfigure\figure
\let\endorigfigure\endfigure
\renewenvironment{figure}[1][2] {
    \expandafter\origfigure\expandafter[H]
} {
    \endorigfigure
}

\usepackage[numbers]{natbib}

\let\textttOrig=\texttt
\def\texttt#1{\expandafter\textttOrig{\seqsplit{#1}}}
\renewcommand{\seqinsert}{\ifmmode
  \allowbreak
  \else\penalty6000\hspace{0pt plus 0.02em}\fi}

\newlength{\cslhangindent}
\newlength{\csllabelwidth}
 {
  \setlength{\parindent}{0pt}
  \ifodd #1 \everypar{\setlength{\hangindent}{\cslhangindent}}\ignorespaces\fi
  \ifnum #2 > 0
  \setlength{\parskip}{#2\baselineskip}
  \fi
 }%
 {}
\usepackage{calc}

\usepackage[top=3.5cm, bottom=3cm, right=1.5cm, left=2.0cm,
            headheight=2.2cm, reversemp, includemp, marginparwidth=4.5cm]{geometry}

\titleformat{\section}
  {\normalfont\sffamily\Large\bfseries}
  {}{0pt}{}
\titleformat{\subsection}
  {\normalfont\sffamily\large\bfseries}
  {}{0pt}{}
\titleformat{\subsubsection}
  {\normalfont\sffamily\bfseries}
  {}{0pt}{}
\titleformat*{\paragraph}
  {\sffamily\normalsize}

\usepackage{fancyhdr}
\makeatletter
\let\ps@plain\ps@fancy
\fancyheadoffset[L]{4.5cm}
\fancyfootoffset[L]{4.5cm}

\definecolor{linky}{rgb}{0.0, 0.5, 1.0}

\newtcolorbox{repobox}
   {colback=red, colframe=red!75!black,
     boxrule=0.5pt, arc=2pt, left=6pt, right=6pt, top=3pt, bottom=3pt}

\newcommand{\ExternalLink}{%
   \tikz[x=1.2ex, y=1.2ex, baseline=-0.05ex]{%
       \begin{scope}[x=1ex, y=1ex]
           \clip (-0.1,-0.1)
               --++ (-0, 1.2)
               --++ (0.6, 0)
               --++ (0, -0.6)
               --++ (0.6, 0)
               --++ (0, -1);
           \path[draw,
               line width = 0.5,
               rounded corners=0.5]
               (0,0) rectangle (1,1);
       \end{scope}
       \path[draw, line width = 0.5] (0.5, 0.5)
           -- (1, 1);
       \path[draw, line width = 0.5] (0.6, 1)
           -- (1, 1) -- (1, 0.6);
       }
   }

\patchcmd{\@maketitle}{center}{flushleft}{}{}
\patchcmd{\@maketitle}{center}{flushleft}{}{}
\patchcmd{\@maketitle}{\LARGE}{\LARGE\sffamily}{}{}
\def\maketitle{{%
  
  \AB@maketitle}}
\makeatletter
\renewcommand\AB@affilsepx{ \protect\Affilfont}
\renewcommand\AB@affilnote[1]{{\bfseries #1}\hspace{3pt}}
\renewcommand{\affil}[2][]%
   {\newaffiltrue\let\AB@blk@and\AB@pand
      \if\relax#1\relax\def\AB@note{\AB@thenote}\else\def\AB@note{#1}%
        \setcounter{Maxaffil}{0}\fi
        \begingroup
        \let\href=\href@Orig
        \let\texttt=\textttOrig
        \let\protect\@unexpandable@protect
        \def\thanks{\protect\thanks}\def\footnote{\protect\footnote}%
        \@temptokena=\expandafter{\AB@authors}%
        {\def\\{\protect\\\protect\Affilfont}\xdef\AB@temp{#2}}%
         \xdef\AB@authors{\the\@temptokena\AB@las\AB@au@str
         \protect\\[\affilsep]\protect\Affilfont\AB@temp}%
         \gdef\AB@las{}\gdef\AB@au@str{}%
        {\def\\{, \ignorespaces}\xdef\AB@temp{#2}}%
        \@temptokena=\expandafter{\AB@affillist}%
        \xdef\AB@affillist{\the\@temptokena \AB@affilsep
          \AB@affilnote{\AB@note}\protect\Affilfont\AB@temp}%
      \endgroup
       \let\AB@affilsep\AB@affilsepx
}
\makeatother

\renewcommand\Affilfont{\sffamily\small\mdseries}
\ifnum 0\ifxetex 1\fi\ifluatex 1\fi=0 
  \usepackage[T1]{fontenc}
  \usepackage[utf8]{inputenc}

\else 
  \ifxetex
    \usepackage{mathspec}
    \usepackage{fontspec}

  \else
    \usepackage{fontspec}
  \fi
  \defaultfontfeatures{Ligatures=TeX,Scale=MatchLowercase}

\fi
\IfFileExists{upquote.sty}{\usepackage{upquote}}{}
\IfFileExists{microtype.sty}{%
\usepackage{microtype}
\UseMicrotypeSet[protrusion]{basicmath} 
}{}

\usepackage{hyperref}
\hypersetup{unicode=true,
            pdftitle={: The -accelerated Universal Bayesian Imaging Kit},
            pdfborder={0 0 0},
            breaklinks=true}
\usepackage{longtable,booktabs}

\let\addcontentslineOrig=\addcontentsline
\def\addcontentsline#1#2#3{\bgroup
  \let\texttt=\textttOrig\addcontentslineOrig{#1}{#2}{#3}\egroup}
\let\markbothOrig\markboth
\def\markboth#1#2{\bgroup
  \let\texttt=\textttOrig\markbothOrig{#1}{#2}\egroup}
\let\markrightOrig\markright
\def\markright#1{\bgroup
  \let\texttt=\textttOrig\markrightOrig{#1}\egroup}

\usepackage{graphicx,grffile}
\makeatletter
\def\maxwidth{\ifdim\Gin@nat@width>\linewidth\linewidth\else\Gin@nat@width\fi}
\def\maxheight{\ifdim\Gin@nat@height>\textheight\textheight\else\Gin@nat@height\fi}
\makeatother
\setkeys{Gin}{width=\maxwidth,height=\maxheight,keepaspectratio}
\IfFileExists{parskip.sty}{%
\usepackage{parskip}
}{
\setlength{\parindent}{0pt}
\setlength{\parskip}{6pt plus 2pt minus 1pt}
}
\ifx\paragraph\undefined\else
\let\oldparagraph\paragraph
\renewcommand{\paragraph}[1]{\oldparagraph{#1}\mbox{}}
\fi
\ifx\subparagraph\undefined\else
\let\oldsubparagraph\subparagraph
\renewcommand{\subparagraph}[1]{\oldsubparagraph{#1}\mbox{}}
\fi

\title{\texttt{J-UBIK}: The \texttt{JAX}-accelerated Universal Bayesian
Imaging Kit}

        \author[*,1, 2]{Vincent Eberle}
          \author[*,1, 2]{Matteo Guardiani}
          \author[*,1, 2]{Margret Westerkamp}
          \author[1]{Philipp Frank}
          \author[2]{Julian Rüstig}
          \author[1, 3]{Julia Stadler}
          \author[1, 2, 3]{Torsten A. Enßlin}
    	  \affil[*]{These authors contributed equally.}
      \affil[1]{Max Planck Institute for Astrophysics,
Karl-Schwarzschild-Straße 1, 85748 Garching bei München, Germany}
      \affil[2]{Ludwig Maximilian University of Munich,
Geschwister-Scholl-Platz 1, 80539 München, Germany}
      \affil[3]{ORIGINS Excellence Cluster, Boltzmannstr. 2, D-85748
Garching, Germany}
  \date{\vspace{-7ex}}

\begin{document}
\maketitle

\marginpar{

  \begin{flushleft}
  \sffamily\small

  \vspace{2mm}

  {\bfseries Software}
  \begin{itemize}
    \setlength\itemsep{0em}
    \item \href{https://github.com/NIFTy-PPL/J-UBIK}{\color{linky}{Repository}} \ExternalLink
  \end{itemize}

  \vspace{2mm}

  \par\noindent\hrulefill\par

  \vspace{2mm}

  {\bfseries Submitted:} 16.09.2024\\
  {\bfseries Published:} N/A

  \vspace{2mm}
  {\bfseries License}\\
  Authors of papers retain copyright and release the work under a Creative Commons Attribution 4.0 International License (\href{http://creativecommons.org/licenses/by/4.0/}{\color{linky}{CC BY 4.0}}).

  \end{flushleft}
}

\hypertarget{summary}{%
\section{Summary}\label{summary}}

Many advances in astronomy and astrophysics originate from accurate
images of the sky emission across multiple wavelengths. This often
requires reconstructing spatially and spectrally correlated signals
detected from multiple instruments. To facilitate the high-fidelity
imaging of these signals, we introduce the universal Bayesian imaging
kit (UBIK). Specifically, we present \texttt{J-UBIK}, a flexible and
modular implementation leveraging the \texttt{JAX}-accelerated
\texttt{NIFTy.re} \cite{Edenhofer:2024} software as its backend.
\texttt{J-UBIK} streamlines the implementation of the key Bayesian
inference components, providing for all the necessary steps of Bayesian
imaging pipelines. First, it provides adaptable prior models for
different sky realizations. Second, it includes likelihood models
tailored to specific instruments. So far, the package includes three
instruments: Chandra and eROSITA for X-ray observations, and the James
Webb Space Telescope (JWST) for the near- and mid-infrared. The aim is
to expand this set in the future. Third, these models can be integrated
with various inference and optimization schemes, such as maximum a
posteriori estimation and variational inference. Explicit demos show how
to integrate the individual modules into a full analysis pipeline.
Overall, \texttt{J-UBIK} enables efficient generation of high-fidelity
images via Bayesian pipelines that can be tailored to specific research
objectives.

\hypertarget{statement-of-need}{%
\section{Statement of Need}\label{statement-of-need}}

In astrophysical imaging, we often encounter high-dimensional signals
that vary across space, time, and energy. The new generation of
telescopes in astronomy offers exciting opportunities to capture these
signals but also presents significant challenges in extracting the most
information from the resulting data. These challenges include accurately
modeling the instrument's response to the signal, accounting for complex
noise structures, and separating overlapping signals of distinct
physical origin.

Here, we introduce \texttt{J-UBIK}, the \texttt{JAX}-accelerated
Universal Bayesian Imaging Kit, which leverages Bayesian statistics to
reconstruct complex signals. In particular, we envision its application
in the context of multi-instrument data in astronomy and also other
fields such as medical imaging. \texttt{J-UBIK} is built on information
field theory (IFT, \cite{Ensslin:2013}) and the \texttt{NIFTy.re} software
package (Edenhofer et al., 2024), a \texttt{JAX}-accelerated version of
\texttt{NIFTy} (\cite{Selig:2013}; \cite{Steininger:2019}; \citep{Arras:2019}).

Following the \texttt{NIFTy} paradigm, \texttt{J-UBIK} employs a
generative prior model that encodes assumptions about the signal before
incorporating any data, and a likelihood model that describes the
measurements, including the responses of multiple instruments and noise
statistics. Built on \texttt{NIFTy.re}, \texttt{J-UBIK} supports
adaptive and distributed representations of high-dimensional physical
signal fields and accelerates their inference from observational data
using advanced Bayesian algorithms. These include maximum a posteriori
(MAP), Hamiltonian Monte Carlo (HMC), and two variational inference
techniques: metric Gaussian variational inference (MGVI, \cite{Knollmueller:2020} \&
\cite{Ensslin:2013}) and geometric variational inference (geoVI, \cite{Frank:2021}). 
As \texttt{NIFTy.re} is fully implemented in \texttt{JAX},
\texttt{J-UBIK} benefits from accelerated inference through parallel
computing on clusters or GPUs.

Building generative models with \texttt{NIFTy.re} for specific
instruments and applications can be very tedious and labor-intensive.
Here, \texttt{J-UBIK} comes into play which addresses this challenge
from two angles. First, it provides tools to simplify the creation of
new likelihood and prior models and acts as a flexible toolbox. It
implements a variety of generic response functions, such as
spatially-varying point-spread functions (PSFs) (\cite{Eberle:2023})
and enables the user to define diverse correlation structures for
various sky components. Second, \texttt{J-UBIK} includes implementations
for several instruments.

Currently, it supports Chandra, eROSITA pointings, and JWST
observations, with plans to expand this list as the user base grows.
This expansion will provide users with a diverse set of accessible
inference algorithms for various instruments. Ultimately \texttt{J-UBIK}
enables the user, through Bayesian statistics, not only to obtain
posterior samples and hence measures of interest such as the posterior
mean and uncertainty of the signal for a several data sets, but also to
perform multi-instrument reconstructions.

The software has already been applied by \cite{Westerkamp:2023},
and publications on eROSITA pointings and JWST are currently in
preparation. In the future, the set of instruments will be further
expanded to include existing imaging pipelines from \texttt{NIFTy}
and \texttt{NIFTy.re} such as those described in \cite{Platz:2023},
\cite{Roth:2023}, \cite{Hutschenreuter:2023}, as well as new
ones.

\hypertarget{bayesian-imaging-with}{%
\section{\texorpdfstring{Bayesian Imaging with
\texttt{J-UBIK}}{Bayesian Imaging with }}\label{bayesian-imaging-with}}

At the core of the \texttt{J-UBIK} package is Bayes' theorem:

\[ \mathcal{P}(s|d) \propto \mathcal{P}(d|s) \mathcal{P}(s), \]

where the prior \(\mathcal{P}(s)\) represents our knowledge about the
signal \(s\) before observing the data \(d\), and the likelihood
\(\mathcal{P}(d|s)\) describes the measurement process. The posterior
\(\mathcal{P}(s|d)\) is the primary measure of interest in the inference
process. \texttt{J-UBIK}'s main role is to model the prior in a
generative fashion and to facilitate the creation and use of instrument
models to develop the likelihood model. The package includes demos for
Chandra, eROSITA pointings, and JWST, which illustrate how to use or
build these models and how to construct an inference pipeline to obtain
posterior estimates.

\hypertarget{prior-models}{%
\subsection{Prior models}\label{prior-models}}

The package includes a prior model for the sky's brightness distribution
across different wavelengths, which can be customized to meet user needs
in both spatial and spectral dimensions. This model allows for the
generation of spatially uncorrelated point sources or spatially
correlated extended sources, as described by the correlated field model
in \cite{Arras:2022}. In the spectral dimension, the model can be a
power law, describe the correlation structure of the logarithmic flux
using a Wiener process along the spectral axis or combine both of these
models. The prior model's structure is designed to be flexible, allowing
for modifications to accommodate additional dimensions and correlation
structures. Figure \ref{fig:sky} illustrates an example of a simulated
X-ray sky in \texttt{J-UBIK}, sampled from a corresponding generative
prior model with one energy bin. This example features two components:
one representing spatially uncorrelated point sources and the other
representing spatially correlated extended structures. Figure
\ref{fig:sky} shows from left to right the full sky and its components,
the diffuse, extended structures and the point sources.

\begin{longtable}[]{@{}l@{}}
\toprule
Figure 1: Simulated X-ray Sky\tabularnewline
\midrule
\endhead
\includegraphics{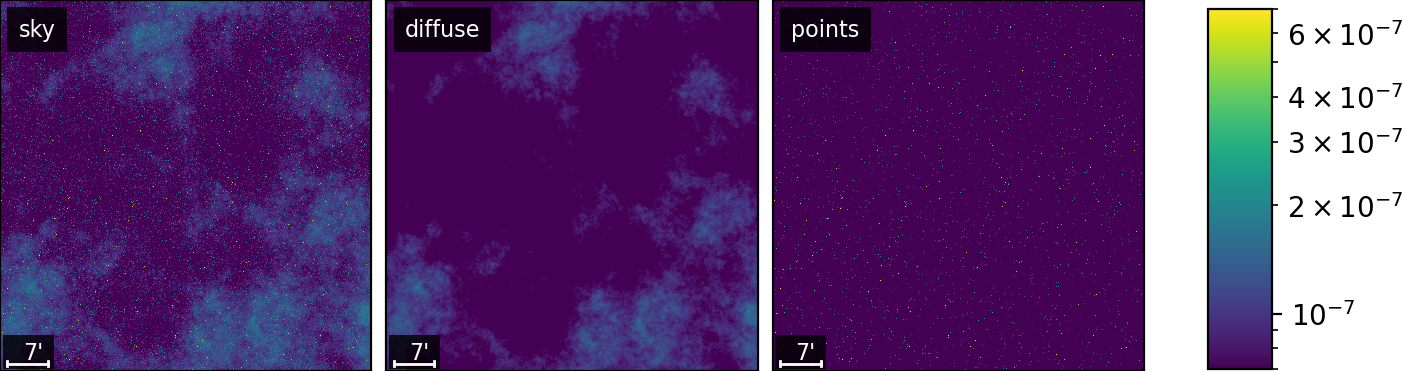}\label{fig:sky}\tabularnewline
\bottomrule
\end{longtable}

\hypertarget{likelihood-models}{%
\subsection{Likelihood models}\label{likelihood-models}}

\texttt{J-UBIK} implements several instrument models (Chandra, eROSITA,
JWST) and their respective data- and response-loading functionalities,
enabling their seamless integration into the inference pipeline. Due to
its fully modular structure, we anticipate the inclusion of more
instruments into the \texttt{J-UBIK} platform in the future.
\texttt{J-UBIK} is not only capable of reconstructing signals from real
data; since each instrument model acts as a digital twin of the
corresponding instrument, it can also be used to generate simulated data
by passing sky prior models through the instrument's response. This
allows to test the consistency of the implemented models.

\begin{longtable}[]{@{}l@{}}
\toprule
Figure 2: Simulated X-ray Data\tabularnewline
\midrule
\endhead
\includegraphics{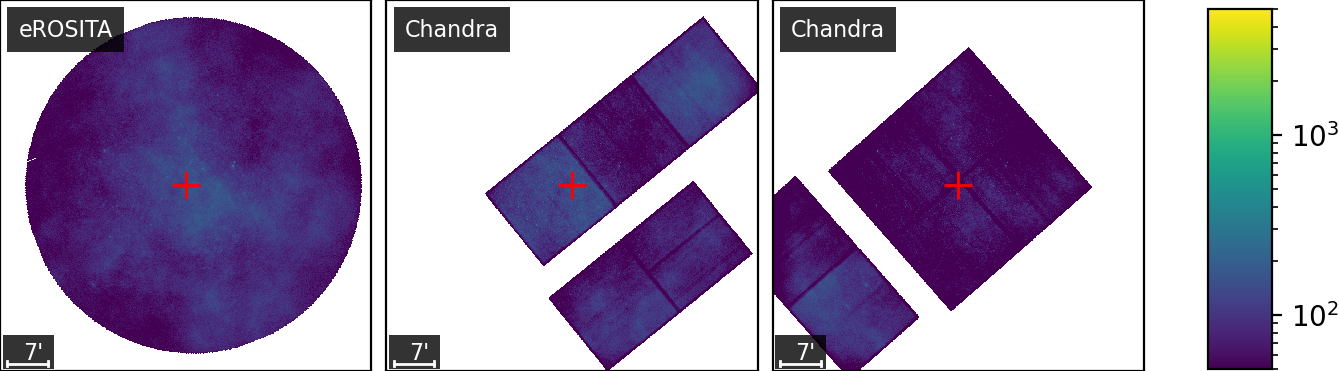}\label{fig:data}\tabularnewline
\bottomrule
\end{longtable}

Figure \ref{fig:data} shows the same simulated sky from Figure
\ref{fig:sky} seen by two different instruments, eROSITA and Chandra,
with Poisson noise on the photon count data. The pointing center for
each observation is marked in red. The two images on the right
illustrate the same simulated sky seen by Chandra, but with different
pointing centers, showing the impact of spatially varying PSFs (\cite{Eberle:2023}).

\hypertarget{acknowledgements}{%
\section{Acknowledgements}\label{acknowledgements}}

V. Eberle, M. Guardiani, and M. Westerkamp acknowledge support for this
research through the project Universal Bayesian Imaging Kit (UBIK,
Förderkennzeichen 50OO2103) funded by the Deutsches Zentrum für Luft-
und Raumfahrt e.V. (DLR). P. Frank acknowledges funding through the
German Federal Ministry of Education and Research for the project
``ErUM-IFT: Informationsfeldtheorie für Experi- mente an
Großforschungsanlagen'' (Förderkennzeichen: 05D23EO1). J. Stadler
acknowledges support by the Deutsche Forschungsgemeinschaft (DFG, German
Research Foundation) under Germany's Excellence Strategy -- EXC-2094 --
390783311.

\bibliographystyle{plainnat}
\bibliography{paper.bib}

\begin{thebibliography}{13}
\providecommand{\natexlab}[1]{#1}
\providecommand{\url}[1]{\texttt{#1}}
\expandafter\ifx\csname urlstyle\endcsname\relax
  \providecommand{\doi}[1]{doi: #1}\else
  \providecommand{\doi}{doi: \begingroup \urlstyle{rm}\Url}\fi

\bibitem[Arras et~al.(2019)Arras, Baltac, En{\ss}lin, Frank, Hutschenreuter,
  Knollmueller, Leike, Newrzella, Platz, Reinecke, et~al.]{Arras:2019}
Philipp Arras, Mihai Baltac, Torsten~A En{\ss}lin, Philipp Frank, Sebastian
  Hutschenreuter, Jakob Knollmueller, Reimar Leike, Max-Niklas Newrzella, Lukas
  Platz, Martin Reinecke, et~al.
\newblock Nifty5: Numerical information field theory v5.
\newblock \emph{Astrophysics Source Code Library}, pages ascl--1903, 2019.

\bibitem[Arras et~al.(2022)Arras, Frank, Haim, Knollmüller, Leike, Reinecke,
  and Enßlin]{Arras:2022}
Philipp Arras, Philipp Frank, Philipp Haim, Jakob Knollmüller, Reimar Leike,
  Martin Reinecke, and Torsten Enßlin.
\newblock Variable structures in m87* from space, time and frequency resolved
  interferometry.
\newblock \emph{Nature Astronomy}, 6\penalty0 (2):\penalty0 259–269, January
  2022.
\newblock ISSN 2397-3366.
\newblock \doi{10.1038/s41550-021-01548-0}.
\newblock URL \url{http://dx.doi.org/10.1038/s41550-021-01548-0}.

\bibitem[Eberle et~al.(2023)Eberle, Frank, Stadler, Streit, and
  Enßlin]{Eberle:2023}
Vincent Eberle, Philipp Frank, Julia Stadler, Silvan Streit, and Torsten
  Enßlin.
\newblock Butterfly transforms for efficient representation of spatially
  variant point spread functions in bayesian imaging.
\newblock \emph{Entropy}, 25\penalty0 (4), 2023.
\newblock ISSN 1099-4300.
\newblock \doi{10.3390/e25040652}.
\newblock URL \url{https://www.mdpi.com/1099-4300/25/4/652}.

\bibitem[Edenhofer et~al.(2024)Edenhofer, Frank, Roth, Leike, Guerdi,
  Scheel-Platz, Guardiani, Eberle, Westerkamp, and Enßlin]{Edenhofer:2024}
Gordian Edenhofer, Philipp Frank, Jakob Roth, Reimar~H. Leike, Massin Guerdi,
  Lukas~I. Scheel-Platz, Matteo Guardiani, Vincent Eberle, Margret Westerkamp,
  and Torsten~A. Enßlin.
\newblock Re-envisioning numerical information field theory (nifty.re): A
  library for gaussian processes and variational inference.
\newblock \emph{Journal of Open Source Software}, 9\penalty0 (98):\penalty0
  6593, June 2024.
\newblock ISSN 2475-9066.
\newblock \doi{10.21105/joss.06593}.
\newblock URL \url{http://dx.doi.org/10.21105/joss.06593}.

\bibitem[Enßlin(2013)]{Ensslin:2013}
Torsten Enßlin.
\newblock Information field theory.
\newblock In \emph{AIP Conference Proceedings}. AIP, 2013.
\newblock \doi{10.1063/1.4819999}.
\newblock URL \url{http://dx.doi.org/10.1063/1.4819999}.

\bibitem[Frank et~al.(2021)Frank, Leike, and En{\ss}lin]{Frank:2021}
Philipp Frank, Reimar Leike, and Torsten~A. En{\ss}lin.
\newblock Geometric variational inference.
\newblock \emph{Entropy}, 23\penalty0 (7):\penalty0 853, jul 2021.
\newblock \doi{10.3390/e23070853}.
\newblock URL \url{https://doi.org/10.3390%2Fe23070853}.

\bibitem[{Hutschenreuter} et~al.(2023){Hutschenreuter}, {Haverkorn}, {Frank},
  {Raycheva}, and {En{\ss}lin}]{Hutschenreuter:2023}
Sebastian {Hutschenreuter}, Marijke {Haverkorn}, Philipp {Frank}, Nergis~C.
  {Raycheva}, and Torsten~A. {En{\ss}lin}.
\newblock {Disentangling the Faraday rotation sky}.
\newblock \emph{arXiv e-prints}, art. arXiv:2304.12350, April 2023.
\newblock \doi{10.48550/arXiv.2304.12350}.

\bibitem[Knollmüller and Enßlin(2020)]{Knollmueller:2020}
Jakob Knollmüller and Torsten~A. Enßlin.
\newblock Metric gaussian variational inference, 2020.

\bibitem[{Roth} et~al.(2023){Roth}, {Arras}, {Reinecke}, {Perley},
  {Westermann}, and {En{\ss}lin}]{Roth:2023}
Jakob {Roth}, Philipp {Arras}, Martin {Reinecke}, Richard~A. {Perley},
  R{\"u}diger {Westermann}, and Torsten~A. {En{\ss}lin}.
\newblock {Bayesian radio interferometric imaging with direction-dependent
  calibration}.
\newblock \emph{A \& A}, 678:\penalty0 A177, October 2023.
\newblock \doi{10.1051/0004-6361/202346851}.

\bibitem[Scheel-Platz et~al.(2023)Scheel-Platz, Knollmüller, Arras, Frank,
  Reinecke, Jüstel, and Enßlin]{Platz:2023}
L.~I. Scheel-Platz, J.~Knollmüller, P.~Arras, P.~Frank, M.~Reinecke,
  D.~Jüstel, and T.~A. Enßlin.
\newblock Multicomponent imaging of the fermi gamma-ray sky in the
  spatio-spectral domain.
\newblock \emph{Astronomy \& Astrophysics}, 680:\penalty0 A2, December 2023.
\newblock ISSN 1432-0746.
\newblock \doi{10.1051/0004-6361/202243819}.
\newblock URL \url{http://dx.doi.org/10.1051/0004-6361/202243819}.

\bibitem[{Selig} et~al.(2013){Selig}, {Bell}, {Junklewitz}, {Oppermann},
  {Reinecke}, {Greiner}, {Pachajoa}, and {En{\ss}lin}]{Selig:2013}
M.~{Selig}, M.~R. {Bell}, H.~{Junklewitz}, N.~{Oppermann}, M.~{Reinecke},
  M.~{Greiner}, C.~{Pachajoa}, and T.~A. {En{\ss}lin}.
\newblock {NIFTY - Numerical Information Field Theory. A versatile PYTHON
  library for signal inference}.
\newblock \emph{A \& A}, 554:\penalty0 A26, June 2013.
\newblock \doi{10.1051/0004-6361/201321236}.

\bibitem[{Steininger} et~al.(2019){Steininger}, {Dixit}, {Frank}, {Greiner},
  {Hutschenreuter}, {Knollm{\"u}ller}, {Leike}, {Porqueres}, {Pumpe},
  {Reinecke}, {{\v{S}}raml}, {Varady}, and {En{\ss}lin}]{Steininger:2019}
Theo {Steininger}, Jait {Dixit}, Philipp {Frank}, Maksim {Greiner}, Sebastian
  {Hutschenreuter}, Jakob {Knollm{\"u}ller}, Reimar {Leike}, Natalia
  {Porqueres}, Daniel {Pumpe}, Martin {Reinecke}, Matev{\v{z}} {{\v{S}}raml},
  Csongor {Varady}, and Torsten {En{\ss}lin}.
\newblock {NIFTy 3 - Numerical Information Field Theory: A Python Framework for
  Multicomponent Signal Inference on HPC Clusters}.
\newblock \emph{Annalen der Physik}, 531\penalty0 (3):\penalty0 1800290, March
  2019.
\newblock \doi{10.1002/andp.201800290}.

\bibitem[{Westerkamp, M.} et~al.(2024){Westerkamp, M.}, {Eberle, V.},
  {Guardiani, M.}, {Frank, P.}, {Scheel-Platz, L.}, {Arras, P.}, {Knollmüller,
  J.}, {Stadler, J.}, and {Enßlin, T.}]{Westerkamp:2023}
{Westerkamp, M.}, {Eberle, V.}, {Guardiani, M.}, {Frank, P.}, {Scheel-Platz,
  L.}, {Arras, P.}, {Knollmüller, J.}, {Stadler, J.}, and {Enßlin, T.}
\newblock The first spatio-spectral bayesian imaging of sn1006 in x-rays.
\newblock \emph{A\&A}, 684:\penalty0 A155, 2024.
\newblock \doi{10.1051/0004-6361/202347750}.
\newblock URL \url{https://doi.org/10.1051/0004-6361/202347750}.

\end{thebibliography}

\end{document}